\documentclass[runningheads,letterpaper,11pt]{llncs}
\usepackage[hmargin=1.45in,vmargin=1.45in]{geometry}

\usepackage{amssymb,amsmath,latexsym}
\usepackage{graphicx}
\usepackage[font=small,skip=0pt]{caption}
\graphicspath{{images/}}
\usepackage{verbatim}
\usepackage{xcolor}
\usepackage{algorithm,algorithmic}
\usepackage{float}
\usepackage{cite}
\usepackage{autoaligne}
\usepackage{relsize}
\usepackage{url}
\usepackage{hyperref}
\usepackage{varwidth}
\usepackage{footnote}

\usepackage{color,soul}
\soulregister\cite7
\soulregister\ref7
\soulregister\pageref7

\usepackage{amsthm}

\newtheoremstyle{mydefinitionstyle}
  {\topsep}
  {\topsep}
  {\itshape}
  {0pt}
  {\bfseries}
  {}
  { }
  {\thmname{#1}\thmnumber{ #2}\thmnote{ (#3)}}

\theoremstyle{mydefinitionstyle}

\newtheorem{effect}{Effect}
\newtheorem{mydefinition}{Definition}
\newtheorem{observation}{Observation}
\newtheorem{axiom}{Axiom}
\newtheorem{principle}{Principle}
\newtheorem{myclaim}{Claim}
\newtheorem{hypothesis}{Hypothesis}

\newcommand{\quotes}[1]{``#1''} 

\urldef{\mailsa}\path|wiseman.m2500@gmail.com|

\newcommand{\keywords}[1]{\par\addvspace\baselineskip
\noindent\keywordname\enspace\ignorespaces#1}

\usepackage{fancyhdr}
\fancypagestyle{plain}{%
   \fancyhf{}
\fancyfoot[C]{\scriptsize Copyright \textcopyright\ 2019 WizWorld (A Wise World by WizWords). All Rights Reserved.}
   
}

\begin{document}

\mainmatter  

\title{Wise Data: A Novel Approach in Data Science\\from a Network Science Perspective}

\titlerunning{Wise Data: A Novel Approach in Data Science}

\author{Mike Raeini (Mike WiseMan)}
%
\authorrunning{M. WiseMan}

\institute{WizWorld (A Wise World by WizWords) and FAU \\
\href{http://www.WizWorld.net}{www.WizWorld.net}\\
\mailsa\\
}

\maketitle
\thispagestyle{plain}
\pagestyle{headings}

\begin{abstract}
Human beings have been generating data since very long times ago. We ask the following common-sense and wise questions (\textit{WizQuestions}):

\begin{enumerate}
	\item Why do we refer to some \textit{pieces of data} more often than referring to other pieces?
	\vspace{0.1cm}
	\item What does make those commonly-referred pieces of data so \textit{unique} and \textit{different}?
	\vspace{0.1cm}
	\item What are the \textit{characteristics} of data that sometimes make the data so \textit{unique} and \textit{different}?
\end{enumerate}

In this article, we introduce a novel approach (model) that helps us answer these questions from data science and network science perspectives. WizWordily speaking, our proposed approach enables us to model the data (as a network), measure the quality of data, and study the network of data deeply and thoroughly.

\end{abstract}

\section{Introduction}

Human beings have been generating data since a very long time ago. Nowadays, thanks to the electronic revolution, data is generated and dispersed at probably an exponential speed. In fact, the \textit{BuzzWords} \quotes{big data}, \quotes{data science} and \quotes{network science}, that have been around in recent years, were dubbed due to this data revolution. To make a long story short, we have been falling into a vast and deep \textit{ocean of data}, so to speak. This so-called ocean of data, which can also be considered as a very complex \textit{network of pieces of data}, may be studied from different perspectives.
To begin with, one may ask (and ponder about) the following common-sense and wise questions (wizquestions):
\begin{enumerate}
	\item Why do we refer to some \textit{pieces of data} more often than referring to other pieces?
	\item Are all the different pieces of data equally \textit{important}?
	\item What are the \textit{characteristics} of data that sometimes make some pieces of data so \textit{unique} and \textit{different}?
	\item Can we \textit{find} and \textit{extract} the characteristics that make pieces of data different from each other?
	\item Considering the so-called \textit{ocean of data} (the very complex \textit{network of pieces of data}), how does the data \textit{flow} in such an ocean (a network)?
\end{enumerate}
\noindent Some other related questions might be as follows:
\begin{enumerate}
	\setcounter{enumi}{5}
	\item If we want to define the \quotes{\textit{quality}} of data, how can we do that?
	\item Is there a \textit{model (approach)} by which we can \textit{measure} the \textit{quality} of the data that we generate and measure the \textit{quality} of the \textit{works} that we do?
	\item Can we then use such a model \textit{to generate} more quality data, rather than generating data without considering its quality?
	\item Assume we defined a \quotes{\textit{quality}} criteria, how much quality data can then be generated?
	\item Is there a \textit{capacity} (\textit{a limitation}) for the quality data that can be generated?
\end{enumerate}

These questions, among many others, can be interesting to study in data science, network science, in social and behavioral sciences, and even in cognitive science. In this article, we aim at answering these questions from data science and network science perspectives.

\section{The Wise Data (WizData) Model}

In this article, we propose a model that helps us better understand the aforementioned questions. Our model, called WizData, is an approach that enables us to model the data (as a network), measure the quality of data and study the network of data deeply and thoroughly. Data in our model is represented by \textit{pieces of data} and we use the term \textit{word} for \textit{a piece of data}. The criteria that we use for measuring the quality of data is \textit{wiseness}, hence the name Wise Data (WizData). In other words, our model enables us to measure how wise a \textit{word} (a piece of data) is.

First of all, we need to model the data. To do so (and in order to quantize the data), we treat data as pieces of data. We then consider the whole set of pieces of data (i.e. the set of \textit{words}) as a network (mathematical graph), wherein each node is a piece of data (\textit{word}). We use the term \textit{WordNet} to refer to such a network of \textit{words}. A node is connected to another node with a directed arrow if the former node refers to the latter node. For measuring the wiseness of a certain \textit{word} (a certain piece of data), we consider how other \textit{words} in the network judge about the wiseness of that certain \textit{word}. In fact, this approach is borrowed from the way human beings think about wiseness and wisdom.

This approach of modeling data allows us to study the data even more deeply. In fact, when we model the data as a wordnet, we can study the effects of the wordnet on its nodes. In other words, we can study how the whole pieces of data or a subset of pieces of data affect a certain piece of data or a certain subset of pieces of data. In the following sections, we provide the formal definitions of the concepts that we introduce and use in our model. We then propose different approaches for measuring the quality of data. Moreover, we study the network effects on its nodes and provide recommendations for generating quality data. We further discuss the complexity of generating quality data and provide insights on how/where one should look for such data. Finally, we present the observations and results that our model gives us.

We would like to emphasize that the goal of our model is to enable the data generators in a data-driven network to measure the quality of their generated data. This enables the data generators to focus on generating data with higher quality, rather than generating data without considering its quality.

\subsection{Data Visualization with the WizData Model}
\label{subsec:data-visualization}

The WizData model allows us to categorize the data into three different groups or classes. Figure \ref{fig:data-visualization-using-the-wizdata-model} shows how the WizData model categorizes the data:

\begin{figure}[h!]
	\begin{center}
		\includegraphics[width=\textwidth]{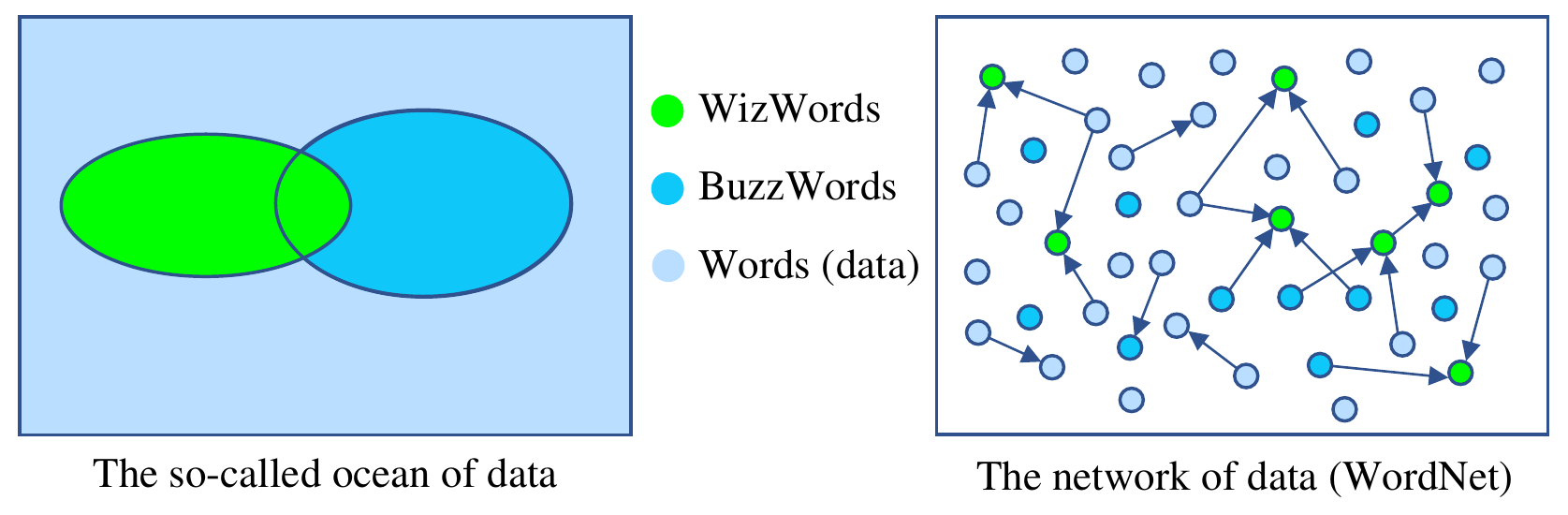}
	\end{center}
	\caption{Data Visualization using the WizData Model}
	\label{fig:data-visualization-using-the-wizdata-model}
\end{figure}

\noindent Note that the network of data is represented by a mathematical directed graph. The nodes of the graph (network) are \textit{pieces of data} (which are called \textit{words}). A directed arrow (edge in the graph) from a node $x_1$ to another node $x_2$ means that node $x_1$ refers to node $x_2$.
Two subnetworks of a wordnet are of particular interest in the WizData model: the network of wizwords and the network of buzzwords.

\begin{figure}[h!]
	\begin{center}
		\includegraphics[width=\textwidth]{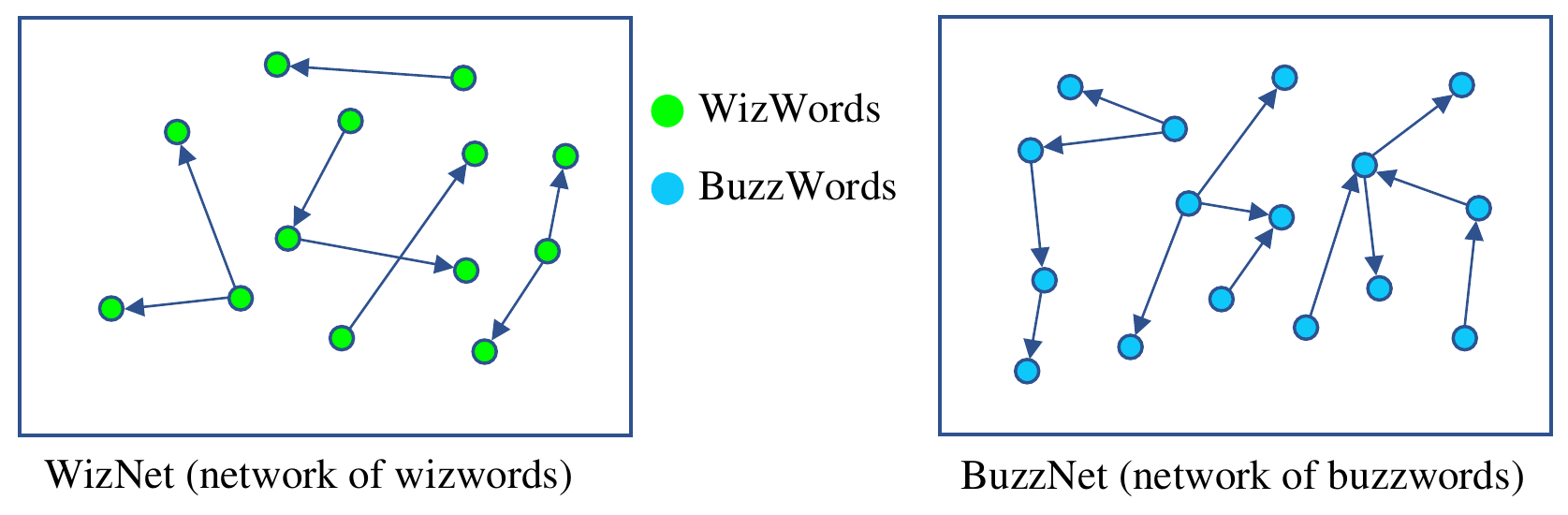}
	\end{center}
	\caption{WizNet (network of wizwords) vs. BuzzNet (network of buzzwords)}
	\label{fig:wiznwt-vs-buzznet}
\end{figure}

\subsection{Definitions: Word, WizScore (WizMeter) and WizWord}
\label{sec:definitions}

In this section, we provide the definition of the concepts that we introduce and use in this article. The most important and key concepts are \textit{Word}, \textit{WizWord} and \textit{WizScore (WizMeter)}. To set everything forth, we should first define what we mean by a \textit{word}.

\begin{mydefinition}{\normalfont \textbf{(\textit{Word}).}}
\label{def:word}
\normalfont A \textit{word} is \textit{a piece of data} in general.
\end{mydefinition}
In the context of our paper, a \textit{word} can be any form of data. More precisely, a \textit{word} can be a word (i.e. a combination of characters), a saying, a piece of text, a statement, a scientific publication (an article, a book etc.), an idea or a thought, a piece of news etc. Regardless of the nature and essence of the data, we treat it as \textit{a piece of data} and we call it a \textit{word}. In general, a \textit{word} may be anything that can be represented on a piece of paper (using a formal or informal language). We emphasize that the term \quotes{\textit{word}} is one of the the key concepts in this paper and we usually use it in \textit{italic} format. Simply, a \textit{word} can be any \textit{piece of data}. Note that the concept of \textit{word}, that we use in the WizData model, allows us to treat data as quantitative entities. It should be mentioned that the length/size of each piece of data does not matter in our model. What matters, however, is the quality of a piece data, regardless of its length or size.
\begin{mydefinition}{\normalfont \textbf{(\textit{\textit{WizScore} or \textit{WizMeter}}).}}
\label{def:wizscore}
\normalfont A \textit{WizScore (WizMeter)} is a mathematical function (metric) that assigns a score or a numeric value to each \textit{word}.
\end{mydefinition}
\begin{mydefinition}{\normalfont \textbf{(\textit{WizWord (wise word or word of wisdom)}).}}
\label{def:wizword}
\normalfont A \textit{WizWord} is a \textit{word} that has a high \textit{wizscore} and also follows the axioms of wizwordiness.
\end{mydefinition}
In Definition \ref{def:wizword}, a \quotes{high \textit{wizscore}} means that a \textit{wizscore} which is higher than a predetermined threshold, e.g., 0.75. Note that in the WizData model, we define \textit{wizscore} to be a numeric value in the $[0, 1]$ interval.
\begin{mydefinition}{\normalfont \textbf{(\textit{WizWork (wise work or work of wisdom)}).}}
\label{def:wizwork}
\normalfont A \textit{WizWork} is a \textit{work} that has a very high \textit{wizscore} and also follows the axioms of wizwordiness.
\end{mydefinition}
\begin{mydefinition}{\normalfont \textbf{(\textit{Value} and \textit{Creating Values}).}}
\label{def:value}
\normalfont A \textit{value} is a valuable (very precious) thing. Creating values means to \textit{create} some \textit{values}.
\end{mydefinition}
In the domain of data, that we deal with in this paper, a \textit{value} can simply be a valuable \textit{word} (piece of data). For instance, a \textit{value} can be a valuable idea, a valuable scientific piece of data etc.
\begin{mydefinition}{\normalfont \textbf{(\textit{Making Differences}).}}
\label{def:making-differences}
\normalfont To change (improve) some things significantly or to create (generate) some new things.
\end{mydefinition}
In the domain of data, making differences means to improve some piece of data significantly, i.e. to increase its quality, or to generate a high quality piece of data.
\begin{mydefinition}{\normalfont \textbf{(\textit{WordNet (network of words)}).}}
\label{def:wordnet}
\normalfont A \textit{WordNet} is a network (graph) of \textit{words}, i.e. a network of pieces of data.
\end{mydefinition}
\begin{mydefinition}{\normalfont \textbf{(\textit{BuzzWord (word of mouth)}).}}
\label{def:buzzword}
\normalfont A \textit{BuzzWord} is a \textit{word} that frequently appears in a wordnet.
\end{mydefinition}
\begin{mydefinition}{\normalfont \textbf{(\textit{WizNet (network of wizwords)}).}}
\label{def:wiznet}
\normalfont A \textit{WizNet} is a network (graph) of \textit{wizwords} or \textit{wizworks}.
\end{mydefinition}
\begin{mydefinition}{\normalfont \textbf{(\textit{BuzzNet (network of buzzwords)}).}}
\label{def:buzznet}
\normalfont A \textit{BuzzNet} is a network (graph) of \textit{buzzwords} or \textit{buzzworks}.
\end{mydefinition}
A very important discussion is that whether a wizword is a true statement or not. In particular, since we define the wizwordiness as a function of other \textit{words}' votes/opinions, one might think that a wizword is always a true statement. We emphasize that in the current WizData model, wizwordiness does not evaluate or measure the truthfulness of a \textit{word}. In fact, determining the truthfulness of a \textit{word} (a piece of data) sometimes does not make that much sense. For instance, in science or mathematics, we cannot talk about the truthfulness of a scientific or mathematical model. However, we can talk about its \textit{factfulness}, its \textit{usefulness}, its \textit{novelty}, and its \textit{quality}.

\begin{figure}[h!]
	\begin{center}
		\includegraphics[width=0.7\textwidth]{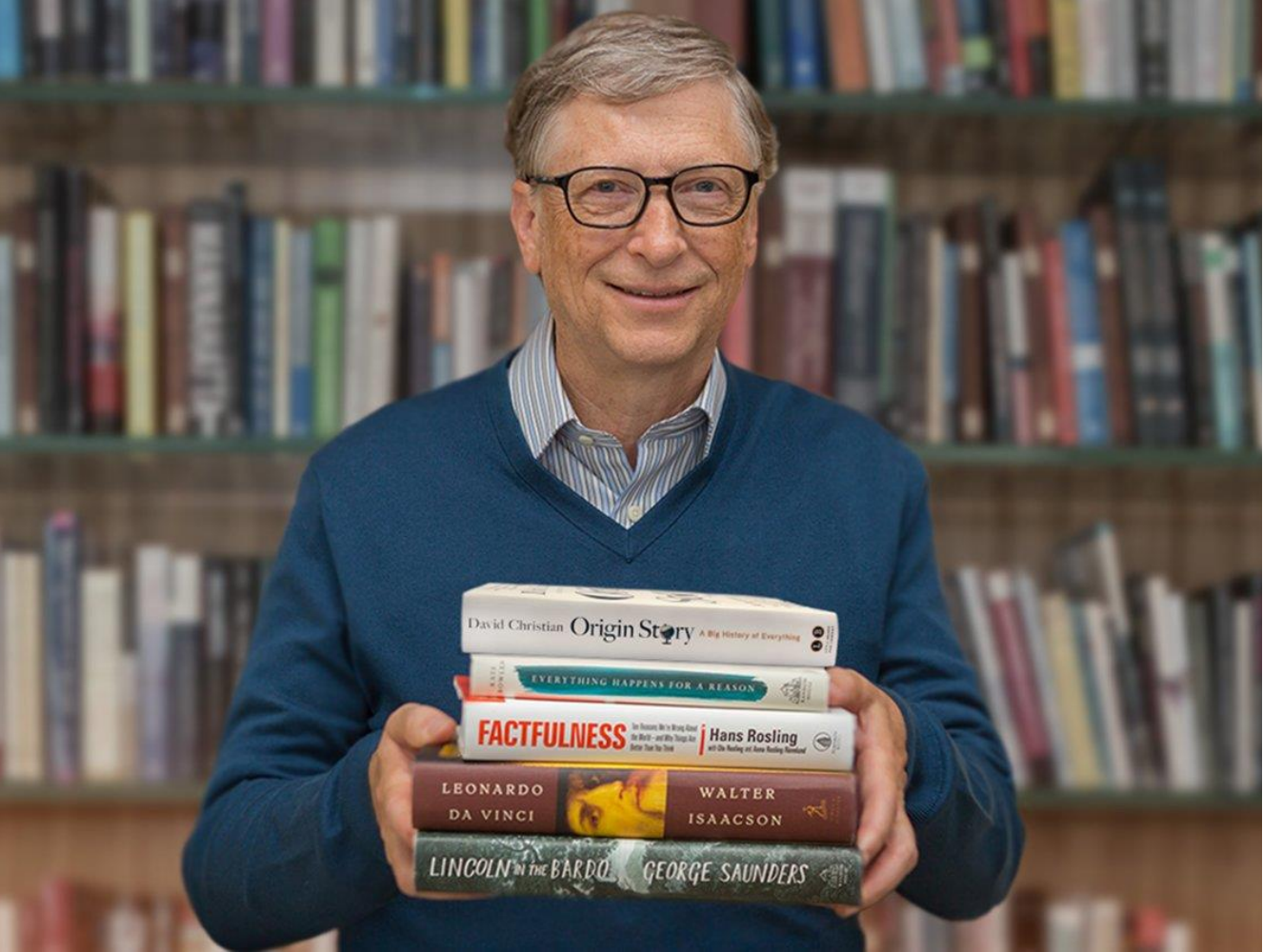}
	\end{center}
	\caption{Factfulness \cite{rosling2019factfulness}, Courtesy of Bill Gates, [Source: BusinessInsider]}
	\label{fig:factfulness}
\end{figure}

Speaking of quality in the domain of data, interested readers may refer to the quality control handbook \cite{juran1951quality}, written by J. M. Juran the pioneer of quality control \cite{juran1992juran} and \cite{juran1975non}. The Pareto principle (a.k.a., the 80/20 rule), which appears in many domains, is claimed \cite{sanders1987pareto} to be named by J. M. Juran (see also \cite{lipovetsky2009pareto}, \cite{wilkinson2006revising}). We will discuss the the emergence of the Pareto principle in complex wordnets in section \ref{sec:observations-results-hypotheses}.

\section{WizScore (WizMeter): a Metric for Measuring the Quality (WizWordiness) of Data}

In this section, we provide some wizscore (wizmeter) functions. Recall that wizscore is a (mathematical) function that assigns a numeric value to a \textit{word} (i.e. a piece of data). The wizscore of a \textit{word }states how wizwordy that \textit{word} is. In other words, wizscore measures the wizwordiness of a \textit{word}, i.e. the quality of a \textit{word} based on the wiseness criteria. Wizscore can be defined in different ways. In what follows, we provide a basic wizscore function and then a more advanced and fair wizscore function. These wizscore functions can be used in different domains. For instance, they can be used for measuring the quality (wizwordiness) of social data (general data in humans' life) or the scientific data.

\subsection{A Basic WizScore (WizMeter)}

We now provide a basic wizscore function. Recall that in the WizData model, data is represented as a network (mathematical graph) of \textit{words} and we use the term \textit{wordnet} for such a network of \textit{words}. Assume a wordnet (network of \textit{words}) is given and let $x$ be a \textit{word} (piece of data) in that wordnet. We define the wizscore of \textit{word} $x$ based on the number of other \textit{words} that refer to \textit{word} $x$. For this purpose and to define a normalized wizscore, we assume that $Max\_Ref$ is the maximum among all the number of references to the \textit{words} in the network. Now, the wizscore of \textit{word} $x$, $WizScore(x)$, is defined as follows:
\begin{equation}
WizScore(x) = \frac{\textrm{The number of references to \textit{word} $x$}}{Max\_Ref}
\end{equation}
It is evident that wizscore is a real number in the interval [0, 1]. We can also define wizscore as an integer number as follows:
\begin{equation}
WizScore(x) =  \lfloor WizScore(x) * 100 \rceil \%
\end{equation}
where $\lfloor a\rceil$, for an arbitrary number $a$, means the closet integer number to $a$. In this case, wizscore is an integer number in $\lbrace 0, 1, 2, ..., 98, 99, 100 \rbrace$.
\vspace{0.2cm}\\
\noindent \textbf{Example: } Assume a wordnet (network of \textit{words}) is given and let the maximum number of references in the network be 1000, i.e., $Max\_Ref = 1000$. Now, we want to measure the wizwordiness of a \textit{word} $x$, which has been referenced 100 times. In this case, the wizscore of $x$ is calculated as follows:
\begin{equation*}
WizScore(x) = \frac{\textrm{The number of references to $x$}}{Max\_Ref} = \frac{100}{1000} = 0.1
\end{equation*}
The wizscore percentage of $x$ would be:
\begin{equation*}
WizScore(x) = \lfloor0.1 \times 100 \rceil = 10
\end{equation*}
If the number of references to $x$ was, for example, 900, then we have:
\begin{equation*}
WizScore(x) = \frac{\textrm{The number of references to $x$}}{Max\_Ref} = \frac{900}{1000} = 0.9
\end{equation*}
The wizscore percentage of $x$ would be:
\begin{equation*}
WizScore(x) = \lfloor 0.9 \times 100 \rceil = 90
\end{equation*}
This means that in the second case, the \textit{word} $x$ is more wizwordy than the \textit{word} $x$ in the first case. In other words, the wizwodiness (quality) of the \textit{word} $x$ in the second case is more/larger than the wizwodiness of $x$ in the first case.

\subsection{A more Advanced and Fair WizScore (WizMeter)}

The basic wizscore we defined above is very simple. Although it may work quite well in some domains, it may not always capture the wizwordiness of a \textit{word}. A drawback of the basic wizscore, as defined above, is that it gives the same weights to all the \textit{words} that refer to the \textit{word} $x$. In fact, the weight for each \textit{word} that refers to the \textit{word} $x$ is one. In order to define a more fair wizscore, we can define a weighted wizscore, wherein each referrer's wizscore is also considered.

Similar to the basic wizscore, assume a wordnet (network of \textit{words}) is given and let $x$ be a \textit{word} in the given wordnet. We want to define and calculate the wizscore of the \textit{word} $x$. Suppose $n$ other \textit{words} $x_1$, $x_2$, $x_3$, ..., $x_n$ in the wordnet refer to the \textit{word} $x$. Moreover, let $Max\_Ref$ be the maximum number of references in the given wordnet. We define a weighted normalized wizscore function as follows (in which the wizscore of each referrer, i.e. $WizScore(x_i)$, is also considered):
\begin{equation}
\label{eq:fair-wizscore}
WizScore (x) = \frac{\sum_{i = 1}^n WizScore(x_i) \times WizScore_{x_i}(x)}{Max\_Ref}
\end{equation}
wherein $WizScore (x_i)$ is the wizscore of referrer $i$, i.e. \textit{word} $x_i$, and $WizScore_{x_i}(x)$ is the wizscore that the \textit{word} $x_i$ gives to the \textit{word} $x$.
Since the wizscore function defined in equation \ref{eq:fair-wizscore} is a weighted average of the wizscores that the \textit{words} $x_1$, $x_2$, $x_3$, ..., $x_n$ give to the \textit{word} $x$, this wizscore captures better the notion of wizwordiness.

\section{The Effects of a Complex WordNet on its Nodes}
\label{sec:the-effects-of-wordnets}

The nodes of a very complex wordnet may significantly be affected by the wordnet (as a whole complex and interactive data-driven system). In what follows, we introduce some effects of a complex wordnet on its nodes:
\begin{effect} {\normalfont \textbf{(\textit{The Effect of BuzzWords (Words of Mouth)}).}}
\label{effect:the-effect-of-buzzwords}
\normalfont The nodes of a complex wordnet tend to follow the buzzwords (words of mouth).
\end{effect}
\begin{effect} {\normalfont \textbf{(\textit{The Effect of Follow-the-Flow}).}}
\label{effect:the-effect-of-follow-the-flow}
\normalfont When some flows are formed in a complex wordnet, the nodes (\textit{words}) of the wordnet tend to follow the flows.
\end{effect}
\begin{effect} {\normalfont \textbf{(\textit{The Matthew Effect}).}}
\label{effect:the-matthew-effect}
\normalfont In a complex wordnet, there is a high probability that buzzwords become buzzwordier, while wizwords become rare (more difficult to discover).
\end{effect}
It should be mentioned that the Matthew effect \cite{merton1968matthew} is a well-known concept and it can be observed in many aspects of humans' life, e.g. in economy and sociology.

\section{Axioms of WizWordiness}
\label{sec:axioms-of-wizwordiness}

We can now answer, at least to some extents, the questions that we raised at the beginning of this article. There are different factors (characteristics) that make a \textit{word} (a piece of data) so unique and different (than other \textit{words}). Among other factors, the factfulness of a \textit{word}, its usefulness, its novelty make the \textit{word} unique and different. The \textit{values} that a \textit{word} \textit{creates} and the \textit{differences} that a \textit{word} \textit{makes} are other important factors that help a \textit{word} stand out and on top of other \textit{words}. In fact, these are some of the characteristics that make \textit{words} different. Note that these can be considered as the intrinsic characteristics of a \textit{word}.

In addition, it is very likely that a \textit{word} in a wordnet is affected by the wordnet (as a whole system) or other \textit{words} in the wordnet. Such effects can be considered as external factors or the effects of environment on a \textit{word}. It can be said that when new \textit{words} are generated in a wordnet, at least three main approaches may commonly happen (as the consequences of the effects of the wordnet). These three common approaches are: \quotes{overstepping (going too far)} on previous \textit{words}, \quotes{rehashing} previous \textit{words}, and \quotes{sophisticating} previous \textit{words}. In what follows, we formalize the \textit{axioms of wizwordiness} (axioms of generating quality data) by considering the aforementioned factors that affect the quality of data. These axioms help the data generators in a wordnet to generate \textit{words} with higher quality.
\begin{axiom} {\normalfont \textbf{(\textit{Simplicity (StraightForwardness)}).}}
\normalfont Wizwords are (and should be) straightforward, easy to understand and easy to use.
\end{axiom}
One might claim that this axiom seems to be controversial, because there might be wizwords that are complicated. Although that might be the case in some domains, accepting this axiom does not seem to deteriorate anything. In fact, the results (achievements) of accepting the axiom may outweigh the results of not accepting it. Therefore, it is highly recommended to consider simplicity (straightforwardness) while generating new \textit{words}.
\begin{axiom} {\normalfont \textbf{(\textit{Creating New Values}).}}
\normalfont It is highly recommended to consider \textit{creating (new) values} while generating new \textit{words}.
\end{axiom}
\begin{axiom} {\normalfont \textbf{(\textit{Making Significant Differences}).}}
\normalfont It is highly recommended to consider \textit{making (significant) differences} while generating new \textit{words}.
\end{axiom}
\begin{axiom} {\normalfont \textbf{(\textit{No Overstepping}).}}
\normalfont Overstepping or going too far on a \textit{word} is not recommended.
\end{axiom}
\begin{axiom} {\normalfont \textbf{(\textit{No Rehashing}).}}
\label{axiom:no-rehashing}
\normalfont Rehashing the existing \textit{words} (i.e. the previously generated data) is not recommended.
\end{axiom}
\begin{axiom} {\normalfont \textbf{(\textit{No Sophisticating (Sophistication)}).}}
\normalfont Sophisticating a \textit{word} is not recommended.
\end{axiom}
It can be said that following the above axioms can significantly help the data generators in a wordnet to generate \textit{words} with higher quality.
Generating new wizwords, by definition, results in creating new values and making significant differences.

\section{WizWord Mining, Value Mining and Where the WizWords (thus the Values) are Hidden}

Now, an interesting problem is to understand where the wizwords (and thus the values) are hidden and how one can mine (discover) them. Another interesting question is that how many wizwords can be generated in a wordnet (or in a certain domain). In other words, is there a capacity (limitation) for the amount/number of wizwords that one can generate in a wordnet (or in a certain domain). The importance of studying these questions is that mining (discovering) new wizwords, by definition, results in creating new values and making significant differences. It may also lead to generating new sciences.

Lets start the discussion by reviewing some wizworks and raising some interesting and wise questions (WizQuestions). Nowadays the world relies heavily on electricity, and also on computers, telephone and the Internet, among other things. So, the invention of electricity has been a very important and influential work (a wizwork). It is worth mentioning that Thomas Edison has been known as the wizard of Menlo Park \cite{stross2008wizard}, probably due to his influential works in the area of electrical engineering. The invention of telephone, computers, and the Internet has also been really influential (some other wizworks). It seems such wizworks (i.e. influential works such as the mentioned inventions) happen rarely. Lets ponder about the following interesting and wise questions (wizquestions):
\begin{enumerate}
	\item Why do some \textit{wizworks}, e.g. the invention of electricity, happen rarely?
	\item Is that because such \textit{wizworks} are based on some things and those things are rare?
	\item Or, is that because those \textit{wizworks} are based on some things and those things are common or abundant, but difficult to find?
\end{enumerate}

If wizworks (and wizwords) are based on some things and those things are rare, then there is not much for us to do. However, if wizworks are based on some things and those things are common or abundant but difficult to find, then we may find such things by more effort. This discussion depends on the domain (context) that we are talking about. While in science the story may go into physics, chemistry, biology etc., thus probably into the natural and intrinsic properties of matter (material), for mathematics the story seems to be completely different. In what follows, we elaborate on this discussion in the domain of mathematics (which is also related to other domains such as computer science, data science and network science).

\subsection{WizWord Mining in Mathematics (and Computer Science)}
\label{subsec:wizword-mining-in-mathematics}

In mathematics, that we are more familiar with, we claim that wizwords and values are hidden in the properties of mathematical structures and concepts or are based on the power of mathematical models. We provide strong examples to support our claim. To begin with, recall that we already defined the concept of \textit{value} (definition \ref{def:value} in section \ref{sec:definitions}). We now need to clarify what we mean by mathematical structures.
For now, lets suffice to some examples. Examples of mathematical structures include: sets, numbers, groups, rings, fields, polynomials, matrices, lattices, graphs, topologies, differential structures, etc. For more details and the formal definition of mathematical structures, interested readers may refer to \cite{corry1992nicolas} and \cite{hegedus2011emergence}, among other references.

First of all, it is worth mentioning that the difference between science and mathematics is in terms of the concepts and objects that are studied. In science (i.e. physics, chemistry, biology etc.), mostly the natural materials are studied; thus values may be created based on the natural properties (characteristics) of the materials. In mathematics, on the other hand, there is no natural material to study. Instead, there are mathematical concepts and structures which are abstract and built in mind. This seems to make the story completely different.

Although natural materials may have limited properties (e.g. chemical, physical or biological characteristics), mathematical concepts and structures might not be limited in terms of the properties that they have. In other words, mathematical concepts and structures may be designed somehow that they have any (number of) desired characteristics that we want. For instance, one can define a new mathematical structure or concept and give certain properties to that structure or concept. We can now assert the following claim and provide strong examples to support it.

\begin{myclaim}{\normalfont\textbf{(\textit{Where the wizwords are hidden}).}}
\label{claim:where-the-wizwords-are-hidden}
\normalfont In mathematics, the wizwords (and thus the values) are hidden in the properties of mathematical structures and concepts, or are based on the power of mathematical models.
\end{myclaim}

A very good example of mathematical structures for supporting the above claim is the set of matrices. It can be said that matrices is one of the very powerful and mostly used mathematical structures. Examples of their applications include: in image representation and image processing, in artificial neural networks, in dimensionality reduction for high-dimensional data, in vector spaces, in lattices, in physics and quantum computing, among many other applications. A very powerful property of matrices is that they have eigenvalues and eigenvectors. This nice property of matrices has been used in different applications. In particular, it seems with the group of square matrices one can define a full homomorphism (that preserve eigenvalues and eigenvectors). Such a full homomorphism can be used in fully homomorphic encryption schemes, which allows computation on encrypted data.

A much more interesting thing seems to be that matrices were not defined and some properties (eigenvalues and eigenvectors, their decomposition etc.) were purposely given to them. It seems that it is most likely the other way around. That is, matrices were defined and they happened to have some properties. Then those properties were discovered, maybe many years after developing matrices as a mathematical structure. However, to see how the story of matrices has been going on (and more importantly what other properties they have) may need more deep research and studies. Interested readers may refer to the history of mathematics literature, e.g. \cite{katz2008history}, which also covers the history of matrices.

Another very useful mathematical structure is the set of polynomials. Polynomials have also some appealing properties. For instance, by having sufficient number of points on a polynomial, that polynomial can be reconstructed (calculated). This property can be very helpful: if one has certain number of points on a polynomial and some of the points change (e.g. due to errors), the polynomial can still be reconstructed. This characteristic of polynomials was elegantly used in Reed-Solomon error correcting codes \cite{reed1960polynomial} and their decoding algorithm, known as Berlekamp-Welch decoding algorithm \cite{welch1986error}.

Graphs (and trees) is another well-known and strong mathematical structure. This mathematical structure has been used in different application domains. A very recent application of this mathematical structure is in  Bitcoin and Blockchain (which have been booming since their inception). Blockchain (and also Bitcoin) is based on a specific kind of mathematical graph (more specifically tree) called hash tree or Merkle tree \cite{merkle1982method}. The concept of hash tree was introduced by Ralph Merkle in 1980s \cite{merkle1982method}. Graphs and trees not only were used in the recent technology of blockchain and Bitcoin, they have also been utilized quite vastly in other applications, e.g. in network science, in data structures, and in decision trees as a data mining technique.

Having said all that in support of claim \ref{claim:where-the-wizwords-are-hidden}, the questions of how much (in fact how many) wizwords can we generate, and is there a capacity to the amount of wizwords that we can generate boil down to the following questions. Is there any properties of the existing mathematical structures and concepts that have not been used?
Can we create new mathematical structures, concepts and models, and give them some specific properties that we desire? This story may go into the history of science and the history of mathematics. It may also go into cognitive science, as how powerful human mind is (e.g. for generating new ideas).

\subsection{The Flow of \textit{Words} in a WordNet}

A very interesting area of study in a complex wordnet is to see how the \textit{words} (pieces of data) flow in such a data-driven network. Such a study may be very helpful in different ways. For one, it can help us in generating new wizwords, which by definition results in creating new values and making significant differences. Second, it can be helpful in assessing the effects of \textit{words} on each other, e.g. whether the \textit{words} affect each other positively or negatively. This, in turn, has its own benefits. For instance, one may discuss further about wordnet optimization (considering the wordnet as a complex and very interactive data-driven system) and how to reduce the effect of \textit{words} that negatively affect other \textit{words}. In this section, we elaborate on this discussion by providing some examples from mathematics and computer science.

The wordnet that humans have generated so far is really enormous and complex. When one studies and looks through this wordnet carefully, they may make interesting observations. Lets provide some examples in a very small portion of this wordnet. Numbers (natural numbers and prime numbers) is probably one of the earliest mathematical structures or concepts that human beings have developed. Numbers have very interesting and beautiful properties. Some of the properties of natural numbers have been elegantly used in different applications. For instance, the two breakthrough works (the RSA algorithm and the Diffie-Hellman algorithm), which were introduced in around 1980s, are based on natural numbers and one-way functions, which is another mathematical concept.

The interesting thing seems to be that shortly after those discoveries (the RSA algorithm and the Diffie-Hellman algorithm) based on natural numbers, scientists and researchers tried to discover similar properties in other mathematical structures. For example, it has been found some one-way functions in elliptic curves and more interestingly in lattices, which both are mathematical structures. This shift from natural numbers to other mathematical structures seems to be interesting to study. First of all, it should be mentioned that when early humans were developing natural numbers, those people most likely did not have any idea about using them in modern days technologies, e.g. in computers, the Internet, telecommunication etc.

Considering there has been some sorts of flow (move or shift) of ideas from natural numbers to other mathematical structures (e.g. elliptic curves or lattices), an interesting and wise question is: Why does that happen? That is, why there has been a shift or move of ideas from natural numbers to other mathematical structures? Note that such a move (shift) may also be observed in many other domains. The WizData model can answer this question, at least to some extent. Some very likely possibilities for such flow of ideas might be for creating new values and making significant differences. Of course, curiosity is in human' nature and they have to live their lives.

\subsection{WizPath: The Path from a WizWord to a WizWork}

In a complex wordnet, there are different interesting things which might be worth studying. First, an interesting question is whether there is some main structure inside a very complex wordnet (in the sense that the wordnet is based on and grows on that main structure). The answer to this question is most likely the concept of wiznet. Recall that a wiznet is a network of wizwords and/or wizworks (definition \ref{def:wiznet} in section \ref{sec:definitions}). Another interesting thing in a complex wordnet is to study the paths from wizwords to wizworks (or from wizwords to wizwords). It can be said that such paths play an important role in the life and evolution of a wordnet. We use the term \textit{WizPath} for a path from a wizword to a wizwork.

To better understand the concept of wizpath and its importance, we provide some examples of wizpaths. In section \ref{subsec:wizword-mining-in-mathematics}, we discussed how graphs and polynomials (as mathematical structures) were used in generating wizwords in mathematics. Now considering graphs as a wizword, Merkle tree \cite{merkle1982method} as another wizword and the idea of currency in economics (in particular the idea of cryptocurrency) as a third wizword, it is interesting to see how a path has been established among these wizwords. Another wizpath is the path from polynomials, to the Reed-Solomon error-correcting codes and to their decoding algorithm, as discussed in section \ref{subsec:wizword-mining-in-mathematics}. An interesting wizpath in the domain of science is the path from the early \textit{words} in physics research which ended at the wizworks of the wizard of Menlo Park (Thomas Edison) \cite{stross2008wizard}.

Having that said, there are other things in a wordnet that might seem interesting for further study, i.e., how to mine (discover) wizpaths and how to predict possible future wizpaths. Although at a high level the wizpaths of a wordnet can be easily visible, as a wordnet grows and becomes more complex, at some levels it will be very difficult to find the wizpaths in the wordnet. Finding and predicting wizpaths in a complex wordnet are important in the sense that they can lead to generating new wizwords, which by definition can result in creating new values and making significant differences.

\section{WizWord Complexity (the likelihood of wizwording)}

In this section, we introduce the concept of wizowrd complexity in a wordnet, i.e. how difficult (likely) it is to  generate a wizword (to wizword) in a wordnet. Recall that a wordnet is a network of pieces of data, i.e. a network of \textit{words}.

We define two types of wizword complexity, i.e. global wizword complexity and local wizword complexity. First, we define the complexity of generating a completely brand-new wizword in a wordnet. Such a wizword can be considered as a global wizword in the whole wordnet. Assume a wordnet is given and let the number of already generated wizwords in the wordnet be $n$. Then, we define the complexity (likelihood) of generating a new global wizword as follows:

\begin{equation}
\label{eq:wizword-complexity-global}
Likelihood(wizword_{new}) = \frac{1}{n + 2}
\end{equation}

Note that in equation \ref{eq:wizword-complexity-global}, we added 2 to the denominator (i.e. $n + 2$), for smoothing. This is because sometimes $n$ might be 0, e.g. in a wordnet wherein no wizword has been discovered. In such a case, by adding 2 as the smoothing factor, the complexity of generating a new wizword would be $\frac{1}{2}$, which makes sense.

Second, we define the complexity of generating a new wizword based on an existing wizword in a wordnet. This type of wizword can be considered as a local wizword, i.e. a wizword which is based on another existing wizword. Assume $x$ is an already generated wizword in a wordnet and let the number of already generated wizwords based on $x$ be $m$. We define the complexity (likelihood) of generating a new local wizword based on $x$ as follows:

\begin{equation}
Likelihood(wizword_{new}|x) = \frac{1}{m + 2}
\end{equation}

Note that the two above complexities are not comparable, because they measure two different criteria, i.e. local wizword complexity and global wizword complexity. In fact, the defined complexities give us an estimation of how difficult (likely) it is to generate a new wizword in a wordnet.

\section{Observations, Results and Hypotheses}
\label{sec:observations-results-hypotheses}

The Wise Data (WizData) model led us to interesting observations and results. In this section, we present the observations that we made using our model.
The first observation is an obvious but very important one, which is worth highlighting. We state it in the form of a principle as follows:
\begin{principle} {\normalfont \textbf{(\textit{The Principle of Repetition}).}}
\normalfont The \textit{words} in a complex wordnet are repeated in different ways.
\end{principle}
This principle has, in fact, different important messages. First, although \textit{words} are repeated, they may have different qualities. \textit{Words}, in general, are a combination of the letters of some alphabet (of a formal or informal language). So, what does make the \textit{words} different from each other? The answer is their wizwordiness (i.e. the factfulness and usefulness of the \textit{words}, the novelty behind the \textit{words}, the values they create and the differences they make). The second message of the above principle is that the \textit{words} in a complex wordnet tend to be repeated. This can be due to the effect of complex wordnets on their nodes. Finally the principle conveys another important message. Even repeating \textit{words} may sometimes result in new wizwords, e.g. by injecting/inducing some novelty to the existing \textit{words} or by using the existing \textit{words} in different application domains.
\begin{observation} {\normalfont \textbf{(\textit{The fair WizScore vs. the basic WizScore}).}}
\label{observation:fair-wizscore-vs-simple-wizscore}
\normalfont The fair wizscore works better than the basic wizscore in differentiating between the \textit{words}, i.e. between wizwords and buzzwords.
\end{observation}
\begin{observation} {\normalfont \textbf{(\textit{WizWord vs. BuzzWord}).}}
\label{observation:wizword-vs-buzzword}
\normalfont A wizword may be worth (in terms of value and quality) hundreds, thousands or even millions of \textit{words} or buzzwords.
\end{observation}
\begin{observation} {\normalfont \textbf{(\textit{Difficulty of Wizwording}).}}
\label{observation:difficulty-of-wizwording}
\normalfont Generating wizwords in some of parts of a complex wordnet may need tens, hundreds or thousands times the amount of time and resources needed to generate a wizword in some other parts of the wordnet.
\end{observation}
\begin{hypothesis}{\normalfont\textbf{(\textit{The Pareto Principle in a Wordnet}).}}
\label{hypothesis:the-proportion-of-wizwords-to-words-in-a-wordnet}
\normalfont The proportion of the number of wizwords to the number of the whole \textit{words} in a complex wordnet most likely follows the Pareto principle.
\end{hypothesis}

Note that Hypothesis \ref{hypothesis:the-proportion-of-wizwords-to-words-in-a-wordnet} is based on the Pareto principle \cite{pareto1964cours}, a.k.a. the 80/20 rule or the law of the vital few \cite{juran1975non}. The Pareto principle is observed in many domains in real world, e.g., in economics, business and computer science.
\begin{hypothesis}{\normalfont\textbf{(\textit{The Power-Law Distribution in a Wordnet}).}}
\label{hypothesis:the-power-law-distribution-in-a-wordnet}
\normalfont The wizwordiness of the \textit{words} in a complex wordnet most likely follows the power-law distribution.
\end{hypothesis}

\begin{figure}[h!]
	\begin{center}
		\includegraphics[width=0.8\textwidth]{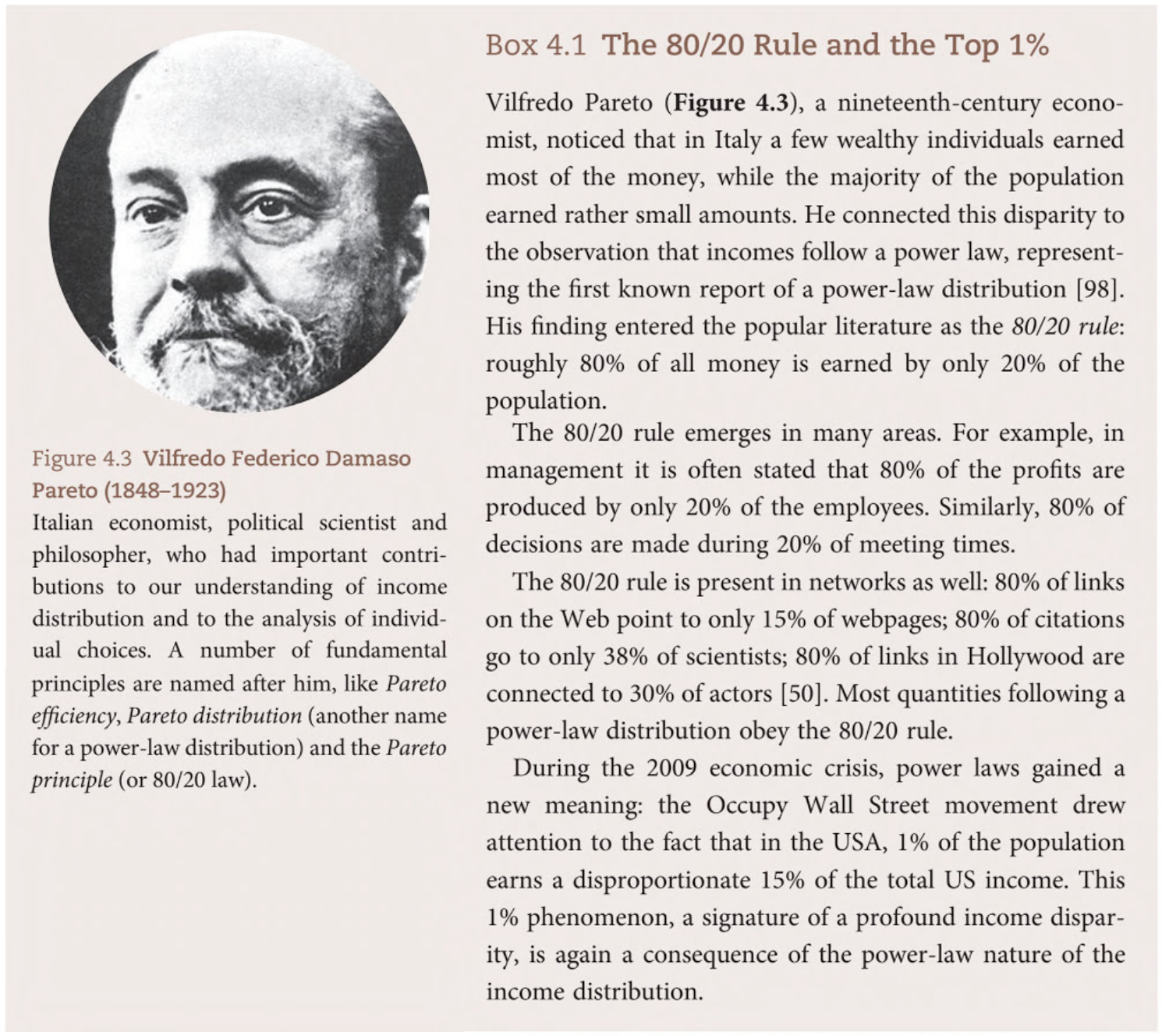}
	\end{center}
	\caption{The 80/20 Rule, Pareto Distribution and Power-Law Distribution  \cite{barabasi2016network}}
	\label{fig:the-80-20-rule}
\end{figure}

Hypothesis \ref{hypothesis:the-power-law-distribution-in-a-wordnet} states that in complex wordnets there are (most likely) many \textit{words} with low wizscores and a few \textit{words} with high wizscores. Note that the power-law distribution and the Pareto principle are very related \cite{barabasi2016network} (see Fig. \ref{fig:the-80-20-rule}).

\begin{hypothesis} {\normalfont \textbf{(\textit{WizWords Matter}).}}
\label{hypothesis:wizwords-matter}
\normalfont Wizwords contribute more to the evolution of a complex wordnet.
\end{hypothesis}
\begin{hypothesis} {\normalfont \textbf{(\textit{The WizNet inside a WordNet}).}}
\label{hypothesis:wiznet-in-wordnets}
\normalfont Inside any complex wordnet there is a simpler wiznet, such that the majority of the \textit{words} of the wordnet are generated based on the wizwords of that wiznet.
\end{hypothesis}
\begin{hypothesis} {\normalfont \textbf{(\textit{The Ratio of Achievements per Consumed Resources}).}}
\label{hypothesis:achievement-per-work-and-resources}
\normalfont The \textit{wizwords} of a complex wordnet most likely bring more values and make more significant differences with a less amount of consumed resources.
\end{hypothesis}
Hypothesis \ref{hypothesis:achievement-per-work-and-resources} can be considered as a consequence of the Pareto principle or the 80/20 rule in complex wordnets \cite{marshall201380}, \cite{koch201380}, and \cite{koch201180}.

\section{Directions for Future Works and Research}

We introduced a novel approach in data science from a network science perspective. Wizwordily speaking, we proposed the WizData model which enables the data generators in a data-driven network to model their data (as a network), to measure the quality of their data and to study the network of their data deeply and thoroughly. We first introduced the concept of \textit{word} as a piece of data. This enabled us to treat the data as quantitative entities (called \textit{words}) and model the data as a network (graph) of \textit{words} (called wordnet). We then introduced some metrics (called WizScore or WizMeter) for measuring the quality of data based on the wiseness criteria. We further discussed the effects of a complex wordnet on its nodes and developed the axioms of wizwordiness. In addition, we discussed where the wizwords (thus the values) are hidden and shed light on where/how one can mine them. Moreover, we introduced the concept of wizword complexity, which is an estimation of how difficult(likely) it is to generate new wizwords.

There are different interesting avenues for further research. The WizData model categorizes the data (\textit{words}) into three groups, i.e. \textit{words}, \textit{wizwords} and \textit{buzzwords}. Recall that figure \ref{fig:data-visualization-using-the-wizdata-model} in section \ref{subsec:data-visualization} shows how different pieces of data (\textit{words}) are categorized into three groups. The WizData model can be used for modeling and categorizing the data in different domains, e.g., the scientific data or the social data (general data in humans' life). This can be helpful in different ways and can potentially lead to interesting findings and results. For instance, it enables the data generators to determine the wizwords of their data, focus more on wizwords and generate \textit{words} with higher qualities, rather than generating \textit{words} without considering their quality. This, by definition, can result in creating new values and making significant differences.

Another line of research is to see how the \textit{words} in a complex wordnet effect each other and how the \textit{words} flow in the wordnet. It can be said that the \textit{words} in a complex wordnet may have different amount of effects on each other. Measuring the effect of \textit{words} on each other and how they affect their whole wordnet can be useful in different ways. For instance, it can be used for finding the \textit{words} that adversely affect other \textit{words} or the whole wordnet. One can then see how the effects of such \textit{words} can be reduced in a wordnet. Measuring the quality and effects of the \textit{words} in a wordnet also paves the way for wordnet optimization. A good criteria for wordnet optimization is to maximize its total wizscore.

Finally, one can use other criteria for measuring the quality of data and take different approaches for studying a wordnet. In the WizData model, we used wiseness as the criteria for measuring the quality of data. One may use other criteria (e.g., factfulness, usefulness, truthfulness, novelty etc.) to measure the quality of data and to study a wordnet.
Note that wiseness is a subjective and contextual-based concept. Therefore, one may define and use other wizscores (different from those that we defined in this article) for measuring the wizwordiness of the data.

For our future works, we are planning to construct wordnets with data from different domains and design experiments to validate (test) our model and our hypotheses.

\section{Conclusion}

In this article, we introduced the Wise Data (WizData) model, which is a novel approach in data science from a network science perspective. Wizwordily speaking, the WizData model allows us to model the data (as a network), to measure the quality of data, and to study the network of data deeply and thoroughly. In the WizData model, data is treated as quantitative entities, called \textit{words}, and modeled as a network of \textit{words}, called wordnet.

The WizData model can help the data generators of a wordnet in different ways. First, the data generators can measure the quality of their data, find the most quality data in their wordnet and focus on generating data with higher quality, rather than generating data without considering its quality. Second, modeling data as a network enables the data generators to study the effects of that network on its nodes, how different pieces of data affect each other, and how the data flows in such a network. Third, a network science approach for studying data helps the data generators to better understand their data and find the relations/connections among different pieces of data. These all can facilitate and accelerate the generation of higher quality data in a wordnet.

The WizData model can be used for modeling and studying data in different domains. In particular, it can be used for modeling the very complex data-driven networks that humans have generated so far, e.g., the scientific data and the humans' social data. Studying such data from a network science perspective can provide us with interesting findings and results.
It may also lead to generating new wizwords, which by definition results in creating new values and making significant differences.

\bibliographystyle{splncs04}
\bibliography{bibliography}

\vspace{1.0cm}

\section*{Note}
Mike Raeini (Mike WiseMan) is also affiliated with the department of CEECS, Florida Atlantic University (FAU), Florida, United States.

\end{document}